\documentclass[fleqn,twoside,twocolumn,nofootinbib,showkeys]{revtex4} % Specifies the document class %,unsortedaddress
\usepackage[nocpr]{ujp} % \usepackage[cyr]{ujp} for cyrillic, \usepackage[web]{ujp} for web
\begin{document}
\title[Nature of the Frequency Shift of Hydrogen Valence Vibrations ]%колонтитул
{NATURE OF THE FREQUENCY SHIFT\\ OF HYDROGEN VALENCE VIBRATIONS\\ IN
WATER MOLECULES}%назва
\author{I.V.~Zhyganiuk}%1 автор
\affiliation{I.I.~Mechnikov National University of Odessa}%институт
\address{2, Dvoryanska Str., Odessa 65026, Ukraine}%адрес
\email{zhyganiuk@gmail.com}%e-mail
\author{M.P.~Malomuzh}%
\affiliation{I.I.~Mechnikov National University of Odessa}%
\address{2, Dvoryanska Str., Odessa 65026, Ukraine}%
\email{zhyganiuk@gmail.com} \udk{???}  \razd{\secix}

\autorcol{I.V.\hspace*{0.7mm}Zhyganiuk, M.P.\hspace*{0.7mm}Malomuzh}

\setcounter{page}{1183}%

\begin{abstract}
The physical nature of a frequency shift of hydrogen valence
vibrations in a water molecule due to its interaction with neighbor
molecules has been studied.\,\,Electrostatic forces connected with
the multipole moments of molecules are supposed to give a dominating
contribution to the intermolecular interaction.\,\,The frequency
shift was calculated in the case where two neighbor molecules form a
dimer.\,\,The obtained result is in qualitative agreement with the
frequency shifts observed for water vapor, hexagonal ice, and liquid
water, as well as for aqueous solutions of alcohols.\,\,This fact
testifies to the electrostatic nature of H-bonds used to describe
both the specific features of the intermolecular interaction in
water and the macroscopic properties of the latter.
\end{abstract}
 \maketitle

\section{Introduction}\raisebox{16mm}[0cm][0cm]{\hspace{8.9cm}\parbox{8.0cm}{{\bfseries\itshape The work is devoted to the 80-th birthday\\
anniversary of the outstanding Ukrainian\\ physicist Galyna
Oleksandrivna Puchkovska}}}

\vspace*{-5mm} \noindent Longitudinal vibrations of the O--H complex
in water molecules and their nearest homologs were studied in plenty
of works carried out during a long-time interval
\cite{1,2,3,4,5,6,7,8}.\,\,Since those vibrations occur along the
line of the chemical O--H bond, they are also often called valence
vibrations.\,\,The frequency of those vibrations in an isolated
water molecule amounts to $\omega _{r}\approx
3657~\mathrm{cm}^{-1}$.\,\,If dimers and, in general, multimers of
higher orders are formed, as well as if water changes its phase
state, the indicated frequency considerably varies \cite{4,7}.
Researches of the frequency shift are an important tool for testing
the various electrostatic models of water molecules \cite{9} and
quantum-mechanical calculations for alcohol multimers \cite{4}.

Note that the frequency shift of valence vibrations in the O--H
complex is observed not only for water molecules at the multimer
formation or the phase state change, but also for alcohol molecules
and in other cases.\,\,In particular, in work \cite{3}, it was shown
that the O--H vibration frequency in the tert-butanol monomer
$\mathrm{C}_{4}\mathrm{H}_{9}\mathrm{-OH}$ equals
4095~\textrm{cm}$^{-1}$, whereas the same frequency in the dimer
$\left( \mathrm{C}_{4}\mathrm{H}_{9}\mathrm{-OH}\right) _{2}$
decreases by 63~$\mathrm{cm}^{-1}$ and equals
4032~\textrm{cm}$^{-1}$.

It is conventionally considered that the frequency shift of
longitudinal vibrations in the O--H complex results from the
formation of a hydrogen bond with the energy $\varepsilon
_{\mathrm{H}}\approx 10\,k_{\rm B}T_{m}$, where $T_{m}$ is the ice
melting temperature.\,\,However, the hydrogen bond concept invokes a
lot of critical remarks \cite{10,11,12,13,14}.\,\,On the one hand,
the hydrogen bond is assumed to have a quantum-mechanical origin
associated with a substantial overlapping of electron shells in the
water molecule \cite{15}.\,\,On the other hand, the thermodynamic
and transport properties of water can be described quite
satisfactorily with the help of intermolecular potentials
\cite{2,9,16,17,18,19,20,21,22,23}.\,\,The latter have the structure
\[
\Phi (q_1 ,q_2 )=\Phi _1^{(1)} (q_1 )+\Phi _1^{(2)} (q_2 )+\Phi
_{\rm int} (q_1 ,q_2 ),\]\vspace*{-7mm}
\[\Phi _{\rm int} (q_1 ,q_2 )=\Phi _{\rm R}
(q_1 ,q_2 )+\Phi _{\rm D} (q_1 ,q_2 )+\Phi _{\rm E} (q_1 ,q_2 ), \]
where $q_{1}$ and $q_{2}$ are the sets of coordinates that describe
the spatial positions and orientations of water molecules; and the
subscripts $R$, $D$, and $E$ denote the repulsion, dispersion, and
electrostatic components, respectively.\,\,The hydrogen bond does
not arise here at all.

%Fig.\,\,1
\begin{figure}%
\vskip1mm
\includegraphics[width=6cm]{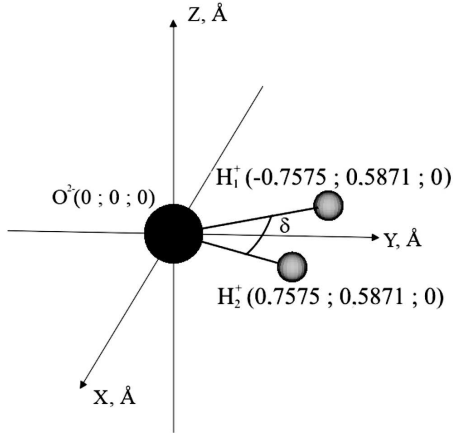}
\vskip-3mm\caption{Electrostatic model of water molecule (the
coordinates of effective charges in Angstr\"{o}m units are indicated
in the parentheses)  }
\end{figure}

In work \cite{24}, it was shown that the hydrogen bond cannot be
ignored completely, since, otherwise, there emerges a difficult
problem concerning the heat capacity of water.\,\,Namely, the latter
is almost twice as high as the heat capacity of hydrogen sulfide,
its nearest homolog.\,\,The solution of this paradox was presented
in works \cite{14,24}.\,\,In particular, the interaction potential
between two water molecules includes also the component
$\Phi_{\mathrm{H}}(q_{1},q_{2})$ corresponding to the energy of the
irreducible hydrogen bond.\,\,Hence,
\[
\Phi _{\rm int} (q_1 ,q_2 )=\Phi _{\rm R} (q_1 ,q_2
)\,+\]\vspace*{-7mm}
\[+\,\Phi _{\rm D} (q_1 ,q_2
)+\Phi _{\rm E} (q_1 ,q_2 )+\Phi _{\rm H} (q_1 ,q_2 ),
\]
At the distances typical of a dimer,
\[
\vert \Phi _{\rm H} (q_1 ,q_2 )\vert \approx \frac{1}{5}\vert \Phi
_{\rm E} (q_1 ,q_2 )\vert ,
\]
i.e. the hydrogen bond is weak and can be taken into account in the
framework of perturbation theory.

The inconsistency is characteristic even of the approach made in
works \cite{9,25} to the problem of the frequency shift of valence
vibrations.\,\,The cited authors proposed one of the soft potentials
describing both the intermolecular interaction and internal
vibrations of water molecules.\,\,However, while calculating the
frequency of valence hydrogen vibrations in a water dimer, a certain
potential of the hydrogen bond is used.\,\,As a result, it was found
that the vibration frequency of hydrogen H$_{1}^{+}$ (see Fig.~2) in
a standard dimer decreases by 300~\textrm{cm}$^{-1}$, which is
considered to prove the existence of a strong hydrogen bond between
water molecules in the dimer.\,\,Note that this shift substantially
exceeds the corresponding value obtained using the methods of IR and
vibrational spectroscopy of water multimers in an argon matrix and
the methods of absorption spectroscopy \cite{7}.

In this work, we give another evidence that irreducible hydrogen
bonds are weak.\,\,We consistently calculated the frequency shift of
longitudinal vibrations of the O--H complex in a water molecule that
forms a dimer configuration with another water molecule.\,\,The
calculations were carried out on the basis of the electrostatic
model of soft water molecule proposed in work \cite{2} and modified
in work \cite{9}.\,\,The interaction between water molecules was
described with the help of the GSD potential \cite{25}.\,\,We showed
that the account of the electrostatic interaction between effective
charges allows the experimentally observed frequency shift for
hydrogen vibrations to be reproduced correctly at the qualitative
level.

\section{Electrostatic Models\\ of Water Molecule and Dimer}

In the initial Stillinger--David model and its generalized GSD
version, the coordinates of three effective charges~-- an oxygen and
two hydrogens~-- are considered variable, unlike the majority of
models \cite{17,18,21,22,23} widely applied in the literature in the
last years.\,\,As a result, the indicated model allows minor
modifications of water molecule parameters to occur under the
influence of the nearest environment.\,\,Let us consider the main
features of the interaction between effective charges in the GSD
model.

\subsection{Electrostatic model of water molecule}

The model of isolated water molecule has a structure shown in
Fig.~1.\,\,We use the potentials in the dimensionless form,
i.e.\,\,the intermolecular interaction potential, as well as other
combinations of constants that have the dimensionality of energy,
are normalized by $k_{\rm B}T_{m}$, where $k_{\rm B}$ is the
Boltzmann constant, and $T_{m}$ the ice melting temperature.\,\,The
corresponding renormalized quantities are tilded.\,\,In particular,
$\tilde{\Phi}=\Phi /k_{\rm B}T_{m}$.

In accordance with the GSD potential \cite{26}, the energy
$\tilde{\Phi}_{1}^{1}$ of the interaction between the hydrogens and the oxygen in a water
molecule $\mathrm{H}_{2}^{+}\mathrm{-O}_{1}^{+}\mathrm{-H}_{1}^{+}$ (see
Fig.~2) is written in the form
%1
\begin{equation}
\label{eq1} {\tilde {\Phi }}_1^{(1)} ( {\tilde {\bf r}}_{{\rm
H}_{1}} , { \tilde {\bf r}}_{{\rm H}_{2} } )=\tilde {\Phi }_{\rm
R}^{(1)} +\tilde {\Phi }_{\rm C}^{(1)} +\tilde {\Phi }_{\rm
Dq}^{(1)} .
\end{equation}
Here, various terms have the following sense:

--~$\tilde{\Phi}_{R}^{1}$ is the potential of the hydrogen repulsion
from the oxygen electron shell:
%2
\begin{equation}
\label{eq2} \tilde {\Phi }_{\rm R}^{(1)} =\tilde {b}_1 \left(\!
{\frac{e^{-\tilde {\rho }_1 {\tilde { r}}_{{\rm H}_{1} } }}{{\tilde
{ r}}_{{\rm H}_{1} } }+\frac{e^{-\tilde {\rho }_1 {\tilde {
r}}_{{\rm H}_{2} } }}{{\tilde { r}}_{{\rm H}_{2} } }} \!\right)\!\!,
\end{equation}
where $\tilde{b}_{1}=b_{1}/k_{\rm B}T_{m}$,
$\tilde{\rho}_{1}=\rho\sigma$, and $\tilde{r}_{{\rm H}_{k}}=r_{{\rm
H}_{k}}/\sigma$ are the amplitude and the reciprocal radius of
repulsion forces between the oxygen and hydrogen atoms, and
$\sigma=2.98$~{\AA} is the diameter of the oxygen atom;

--~$\tilde{\Phi}_{C}^{1}$ is the potential of the direct Coulomb
interaction between the effective charges in the water molecule:
%3
\begin{equation}
\label{eq3} {\tilde {\Phi }}_{\rm C}^{(1)} ={\tilde {C}}_1 \left(\!
\frac{q_{{\rm H}_{1} } q_{{\rm H}_{2} } }{\left| {\tilde {\bf
r}}_{\rm H_{1} } -{\tilde {\bf r}}_{\rm H_{2} }
\right|}+\frac{q_{\rm O_{1}} {q}_{\rm H_{1} } }{{\tilde {r}}_{\rm
H_{1} } }+\frac{q_{\rm O_{1} } q_{\rm H_{2} } }{{\tilde {r}}_{\rm
H_{2} } } \!\right)\!\!,
\end{equation}
where $\tilde{C}_{1}=C_{1}/\sigma=205.59$ is the coefficient of
conversion to dimensionless energy units ($k_{\rm B}T_{m}$), and the
oxygen and hydrogen charges are measured in the electron charge
units and equal $q_{\mathrm{O}_{1}}=-2.0$,
$q_{\mathrm{H}_{1}}=q_{\mathrm{H}_{2}}=1.0$;

--~$\tilde{\Phi}_{Dq}^{1}$ is the potential of the dipole-charge
interaction between the dipole moment of oxygen and the effective
charges of hydrogens:
%4
\[ {\tilde {\Phi }}_{\rm Dq}^{(1)} =\frac{\left( \!{ \tilde {\bf
d}}_{{\rm O}_{1} } {\tilde {\bf r}}_{{\rm H}_{1} }\! \right)q_{{\rm
H}_{1} }}{{\tilde r}_{{\rm H}_{1} } ^3}\left[ 1-{\tilde {L}}({\tilde
{r}}_{{\rm H}_{1} } )\right]+\]\vspace*{-5mm}
\begin{equation}
\label{eq4} +\,\frac{\left( \!{\tilde {\bf d}}_{{\rm O}_{1} } {
\tilde {\bf r}}_{{\rm H}_{2} } \! \right){q}_{{\rm H}_{2} }
}{{\tilde {r}}_{{\rm H}_{2} } ^3}\left[ {1-{\tilde {L}}({\tilde
{r}}_{{\rm H}_{2} }\! )} \right]\!\!,
\end{equation}
where $\tilde{L}(\tilde{r}_{\mathrm{H}_{1}}{)}$ is one of the
screening functions, the explicit form of which is described in work
\cite{26}.\,\,Under the action of hydrogens, the dipole moment of
oxygen in the water molecule acquires the value
%5
 \[{\tilde {\bf d}}_{{\rm O}_{1} } =-{\tilde {\alpha
}}\Biggl\{ \!\frac{{ \tilde {\bf r}}_{{\rm H}_{1} } {q}_{{\rm H}_{1}
} }{{\tilde {r}}_{{\rm H}_{1} } ^3}\left[ {1-{\tilde {K}}({\tilde
{r}}_{{\rm H}_{1} } )} \right]+\]\vspace*{-7mm}
\begin{equation}
\label{eq5} +\frac{{\tilde {\bf r}}_{{\rm H}_{2} } {q}_{{\rm H}_{2}
} }{{\tilde {r}}_{{\rm H}_{2} } ^3}\left[ {1-{\tilde {K}}({\tilde
{r}}_{{\rm H}_{2} } )} \right] \!\!\Biggr\}\!,
\end{equation}

%Fig. 2
\begin{figure}%
\vskip1mm
\includegraphics[width=6cm]{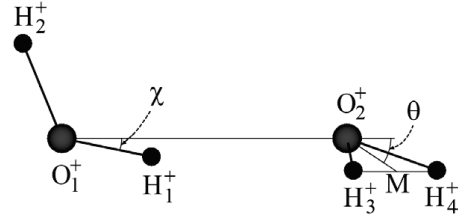}
\vskip-3mm\caption{Configuration of a linear dimer of two water
molecules ($\chi =3.76^{\circ}$ and $\theta=41.1^{\circ}$ are the
equilibrium angle values) }
\end{figure}

\noindent where $\tilde{\alpha}$ is the polarizability of oxygen,
and $\tilde{K}(\tilde{r}_{\mathrm{H}_{2}})$ is another screening
function \cite{2,26}.\,\,The contribution of the dipole-charge
interaction in Eq.\,\,(\ref{eq1}) has the following structure:
%6
\[
{\tilde {\Phi }}_{\rm Dq}^{(1)} =-{\tilde {\alpha }}{\tilde {C}}_1
\Biggl[ {q}_{{\rm H}_{1} } {q}_{{\rm H}_{1} } \frac{\left[
{1-{\tilde {K}}(\tilde r_{{\rm H}_1}\!)} \right]\left[ {1-{\tilde
{L}}(\tilde r_{{\rm H}_1}\!)} \right]}{{\tilde {r}}^6_{{\rm H}_{1} }
}\,\times\]%\vspace*{-7mm}
\[\times\left( { \tilde {\bf r}}_{{\rm H}_{1} } { \tilde {\bf
r}}_{{\rm H}_{1} }  \right)+{q}_{{\rm H}_{1} } {q}_{{\rm H}_{2} }
\,\times
\]\vspace*{-7mm}
\[ \times\frac{\left[
1\!-\!{\tilde {K}}(\tilde r_{{\rm H}_1}\!) \right]\!\left[
1\!-\!{\tilde {L}}(\tilde r_{{\rm H}_2}\!) \right]\!\!+\!\!\left[
1\!-\!{\tilde {K}}(\tilde r_{{\rm H}_2}\!) \right]\!\left[
1\!-\!{\tilde {L}}(\tilde r_{{\rm H}_1}\!) \right]}{{\tilde
{r}}_{{\rm H}_{1} } ^3{\tilde {r}}_{{\rm H}_{2} } ^3}\times
\]\vspace*{-7mm}
\[ \times\left( { \tilde {\bf r}}_{{\rm H}_{1} }  {\tilde {\bf r}}_{{\rm
H}_{2} }  \right)+{q}_{{\rm H}_{2} } {q}_{{\rm H}_{2}
}\,\times\]\vspace*{-7mm}
\begin{equation}
\label{eq6} \times\, \frac{\left[ {1-{\tilde {K}}({\tilde {r}}_{{\rm
H}_{2} } )} \right]\left[ {1-{\tilde {L}}({\tilde {r}}_{{\rm H}_{2}
} )} \right]}{{\tilde {r}}^6_{{\rm H}_{2} } }\left( {\tilde {\bf
r}}_{{\rm H}_{2} }  { \tilde {\bf r}}_{{\rm H}_{2} } \! \right)\!
\Biggr]\!.
\end{equation}
It differs from the expression presented in the work of Stillinger and David
\cite{2} by the coefficient $\frac{1}{2}$.

%Tabl.1
\begin{table*}[!]
\vskip4mm \noindent\caption{Parameters of the GSD potential
}\vskip3mm\tabcolsep12.5pt

\noindent{\footnotesize\begin{tabular}{|c|c|c|c|c|c|c|c|c|c|}
  \hline
  \multicolumn{1}{|c}{\rule{0pt}{5mm}} &
  \multicolumn{1}{|c}{$\tilde {L}_0 $ } &
  \multicolumn{1}{|c}{$\tilde {L}_1 $ } &
  \multicolumn{1}{|c}{$\tilde {L}_2 $ } &
  \multicolumn{1}{|c}{$\tilde {L}_3 $ } &
  \multicolumn{1}{|c}{$\tilde {L}_4 $ } &
  \multicolumn{1}{|c}{$\tilde {b}_1 $ } &
  \multicolumn{1}{|c}{$\tilde {\rho}_1 $ } &
  \multicolumn{1}{|c}{$\tilde {b}_2 $ } &
  \multicolumn{1}{|c|}{$\tilde {\rho}_2 $ } \\[2mm]
  \hline \rule{0pt}{5mm}
  \rule{0pt}{5mm}GSD & 8.8804 & 8.8804&8.1699&124.49&185.95&1064.7&7.656&42129.1&2.59 \\[2mm]
  \hline
\end{tabular}}\vspace*{-3mm}
\end{table*}

\subsection{Interaction of two water molecules}

If one water molecule turns out in the electric field of the other molecule,
its internal energy changes by the magnitude of interaction energy:
%7
\[\tilde {\Phi }_1^{(1)} \!({\tilde {\bf r}}_{{\rm H}_{1} } ,{\tilde {\bf r}}_{{\rm H}_{2} }\! )
\to \tilde {\Phi }_2^{(1)} \!({\tilde {\bf r}}_{{\rm H}_{1} }
,{\tilde {\bf r}}_{{\rm H}_{2} } ,{\tilde {\bf r}}_{{\rm O}_{1} {\rm
H}_{3} } ,{\tilde {\bf r}}_{{\rm O}_{1} {\rm H}_{4} } ,{\tilde {\bf
r}}_{{\rm O}_{1} {\rm O}_{2} }\! )=\]\vspace*{-7mm}
\begin{equation}
\label{eq7} =\tilde {\Phi }_1^{(1)} ({\tilde {\bf r}}_{{\rm H}_{1} }
,{\tilde {\bf r}}_{{\rm H}_{2} } )+\tilde {\Phi }_{\rm Int},
\end{equation}
where the term $\tilde{\Phi}_{\mathrm{Int}}$ looks like
%8
\begin{equation}
\tilde{\Phi}_{\mathrm{Int}}=\tilde{\Phi}_{\rm R}+\tilde{\Phi}_{\rm D}+\tilde{\Phi}%
_{\rm E}.   \label{eq8}
\end{equation}
The first contribution in Eq.~(\ref{eq8}), $\tilde{\Phi}_{R}$, is
the potential of the hydrogen repulsion from the electron shells of
oxygens.\,\,It is approximated by the Born exponential form, which
coincides with a similar contribution in the MPM potential \cite{9}:
%9
\[ \tilde {\Phi }_{\rm R} =\tilde {b}_1 \left[ \sum\limits_{i=1,2}
\frac{e^{-\tilde {\rho }_1 \tilde {r}_{i{\rm O}_{2} } }}{\tilde
{r}_{i{\rm O}_{2} } } +\sum\limits_{j=1,2} \frac{e^{-\tilde {\rho
}_1 \tilde {r}_{{\rm O}_{1} j} }}{\tilde {r}_{{\rm O}_{1} j} }
\right]+\]\vspace*{-5mm}
\begin{equation}
\label{eq9} +\,\frac{\tilde {b}_2 e^{-\tilde {\rho }_2 \tilde
{r}_{{\rm O}_{1} {\rm O}_{2} } }}{\tilde {r}_{{\rm O}_{1} {\rm
O}_{2} } }.
\end{equation}

The second contribution in Eq.~(\ref{eq8}), $\tilde{\Phi}_{\rm D}$,
defines the potential of the dispersion interaction between oxygens:
%10
\begin{equation}
\label{eq10} \tilde {\Phi }_{\rm D} =-\frac{\tilde {A}_{{\rm O}_{1}
{\rm O}_{2} } }{\tilde {r}_{{\rm O}_{1} {\rm O}_{2} }^6 },
\end{equation}
It is similar to the dispersion part of the Lennard-Jones
potential. The same form of the dispersion contribution is inherent to
the SPC and SPC/E potentials \cite{22} and to the TIPS and TIP3P
ones \cite{17}.

The third contribution in Eq.~(\ref{eq8}), $\tilde{\Phi}_{E}$,
defines the electrostatic interaction between oxygens and hydrogens
in two water molecules:
%11
\begin{equation}
\label{eq11} \tilde {\Phi }_{\rm E} =\tilde {\Phi }_{\rm C} +\tilde
{\Phi }_{\rm Dq} +\tilde {\Phi }_{\rm DD},
\end{equation}where the term
%12
\[ \tilde {\Phi }_{\rm C} =\tilde {C}_1 \Biggl\{
\sum\limits_{i,j=1,2}\frac {q_i q_j}{\tilde{r}_{ij}}\,+
\]\vspace*{-5mm}
\begin{equation}
\label{eq12} +\Bigg[\sum\limits_{i=1, 2} \frac {q_i q_{{\rm
O}_2}}{\tilde {r} _{i {\rm O}_2}} + \sum\limits_{j=1,2}
\frac{q_{{\rm O}_1}q_j}{\tilde {r}_{{\rm O}_1 j}}\Bigg]\!+\!
\frac{q_{{\rm O}_1}q_{{\rm O}_2}}{\tilde{r} _{{\rm O}_1{\rm O}_2}}\!
\Biggr\}
\end{equation}
defines the direct Coulomb interaction between the effective charges of two
water molecules and completely coincides with a similar contribution of
the Coulomb interaction in the Stillinger--David potential.

Note that the electrostatic interaction between the effective
charges, which simulate oxygens and hydrogens in the water molecules
in a dimer, consists of two parts: (i)~the interaction of the
effective charges in molecule~1 with the dipole moment of molecule~2
and the interaction of the effective charges in water molecule~2
with the dipole moment of molecule~1, and (ii)~the interaction of
two dipole moments of the water molecules in the dimer.\,\,At
distances between the molecules that considerably exceed the
diameter of a water molecule, the interaction between the model
charges of hydrogens and oxygens completely coincides with the
interaction between the total dipole moments of water molecules.

The structure of the dipole-charge contribution $\tilde{\Phi}_{Dq}$
to the electrostatic interaction is the same as that of the
Stillinger--David potential,
%13
\[ \tilde {\Phi }_{\rm Dq} =\frac{({\tilde {\bf d}}_{{\rm O}_{1} }
{\tilde {\bf r}}_{{\rm O}_{1} {\rm O}_{2} } )q_{{\rm O}_{2} }
}{\tilde {r}_{{\rm O}_{1} {\rm O}_{2} }^3 }[1-\tilde {L}({\tilde
{r}}_{{\rm O}_{1} {\rm O}_{2} } )]\,+\]\vspace*{-7mm}
\[+\,\frac{({\tilde {\bf d}}_{{\rm
O}_{2} }{\tilde {\bf r}}_{{\rm O}_{1} {\rm O}_{2} } )q_{{\rm O}_{1}
} }{\tilde {r}_{{\rm O}_{1} {\rm O}_{2} }^3 }[1-\tilde {L}({\tilde
{r}}_{{\rm O}_{1} {\rm O}_{2} } )]\,+\]\vspace*{-7mm}
\[+\,\Biggl[\, \sum\limits_{j=1,2} \frac{({\tilde {\bf d}}_{{\rm O}_{1} }
{\tilde {\bf r}}_{{\rm O}_{1} j} )q_j }{\tilde {r}_{{\rm O}_{1} j}^3
}[1-\tilde {L}(\tilde {r}_{{\rm O}_{1} j} )] \,+\]\vspace*{-7mm}
\begin{equation}
\label{eq13} +\sum\limits_{i=1,2} \frac{({\tilde {\bf d}}_{{\rm
O}_{2} }{\tilde {\bf r}}_{{\rm O}_{2} i} )q_i }{\tilde {r}_{{\rm
O}_{2} i}^3 }[1-\tilde {L}(\tilde {r}_{{\rm O}_{2} i} )]  \Biggr]\!,
\end{equation}
where the subscripts $i$ and $j$ enumerate the hydrogen charges in
water molecules~1 and 2, respectively; and
$\mathbf{{\tilde{d}}}_{\mathrm{O}_{1}}$ and
$\mathbf{{\tilde{d}}}_{\mathrm{O}_{2}}$ are the dipole moments of
oxygens in the corresponding water molecules.

One can be convinced that, unlike the Stillinger--David potential,
the interaction potential $\tilde{\Phi}_{\mathrm{GSD}}$ has a
correct asymptotics of the dipole-dipole interaction at distances
that strongly exceed the size of a water molecule,
$\tilde{\Phi}_{\mathrm{GSD}}\rightarrow\tilde{\Phi}_{d}({{\tilde{\bf
d}}}_{1}{,}{{{\bf d}}}_{2}),$ where $\tilde {\Phi }_{d} ({\tilde
{\bf d}}_{1} ,{\tilde {\bf d}}_{2} )=$ $=\frac{1}{\tilde {r}_{12}^3
}\left[ \left( {\tilde {\bf d}}_{1} {\tilde {\bf d}}_{1}
\right)-\frac{3\left( {\tilde {\bf d}}_{1} {\tilde {\bf r}}_{12}
\right)\left( {\tilde {\bf d}}_{2} {\tilde {\bf r}}_{21}
\right)}{\tilde {r}_{12}^2 } \right]\!,$ and the dipole moment of
molecule is a sum of the dipole moments of oxygen and hydrogens,
${\tilde{\bf d}}={{\tilde{\bf d}}}_{\mathrm{H}}+{{\tilde{\bf
d}}}_{\mathrm{O}}$. The values of the parameters in the GSD
potential are quoted in Table~1.

Table~2 contains the equilibrium values of the energy
$\tilde{\Phi}_{d}$ (see Fig.~2), the angles $\theta$ and $\chi$, and
the dipole moment $D_{d}$ for the dimer calculated with the use of
the GSD potential.\,\,The direct comparison testifies that they
agree well with the results of quantum chemical calculations
\cite{27} and experimental data \cite{28}.

%Tabl.2
\begin{table*}[!]
\vskip4mm \noindent\caption{Equilibrium values of the distance\\
between the oxygens, the angles, the energy, and the dipole moment
of a water dimer }\vskip3mm\tabcolsep21.6pt

\noindent{\footnotesize\begin{tabular}{|l|c|c|c|c|c|}
  \hline
  \multicolumn{1}{|c}{\rule{0pt}{5mm}Parameters} &
  \multicolumn{1}{|c}{$r_{{\rm O}_1 {\rm O}_2}$, \AA } &
  \multicolumn{1}{|c}{$\gamma$, deg } &
  \multicolumn{1}{|c}{$\theta$, deg } &
  \multicolumn{1}{|c}{$\tilde {\Phi}_d$, \AA } &
  \multicolumn{1}{|c|}{${D}_d$, D } \\[2mm]
  \hline \rule{0pt}{5mm}GSD & 2.96 & 3.76&41.1&--8.01&2.54 \\
  Experiment & $2.976 \pm 0.004$ & $-1\pm 10$&$57\pm 10$&$-9.96\pm 0.4$&2.6~\, \\[2mm]
  \hline
\end{tabular}}\vspace*{-2mm}
\end{table*}

\section{Behavior of the Vibration\\ Frequencies of Hydrogens in a
Water\\
Molecule in Vapor, Water, and Ice}

Let us discuss the shift of longitudinal (valence) vibration
frequencies for the hydrogen $\mathrm{H}_{1}^{+}$ (see Fig.\,\,2)
that lies close to the line connecting the centers-of-mass of oxygen
atoms in two neighbor water molecules forming a dimer.\,\,The
corresponding vibration frequencies are determined according to the
formula
%14
\begin{equation}
\omega _{\parallel}\approx \sqrt{\frac{K_{rr}^{1}}{M_{\rm Re}}},
\label{eq14}
\end{equation}%
where the force constant (in the dimensionless form) is defined in the
standard manner:
%15
\begin{equation}
\label{eq15} \left. \tilde {K}_{rr}^{(1)} =\frac{\partial
^{2}{\tilde {\Phi }}_{2}^{(1)} }{\partial {\tilde {r}}_{{\rm H}_{1}
}^{2} } \right|_ {\tilde { r}'_{{\rm H}_{1} }}  \!\!\!,
\end{equation}
and the reduced mass of the oxygen--hydrogen system in the water
molecule approximately equals
%16
\begin{equation}
\label{eq16} {M}_{\rm Re} \approx \frac{\left( {{M}_{\rm O}
+{M}_{\rm H} } \right){M}_{\rm H} }{\left( {{M}_{\rm O} +{M}_{\rm H}
} \right)+{M}_{\rm H} }.
\end{equation}
The approximate character of formula (\ref{eq16}) is explained by
the fact that hydrogen $\mathrm{H}_{2}^{+}$ (Fig.\,\,2) is not
located on the line connecting the centers-of-mass of
oxygens.\,\,The second-order derivative in Eq.~(\ref{eq15}) is
calculated at the point $\tilde{r}_{\mathrm{H}_{1}}^{\prime }$
defined by the equation
%17
\begin{equation}
\left. \frac{\partial \!\left(\! {\tilde {\Phi }_1^{(1)}\! ({\tilde
{\bf r}}_{{\rm H}_{1} } ,{\tilde {\bf r}}_{{\rm H}_{2} }
)\!+\!\tilde {\Phi }_{\rm Int} ({\tilde {\bf r}}_{{\rm O}_{1} {\rm
H}_{3} } ,{\tilde {\bf r}}_{{\rm O}_{1} {\rm H}_{4} } ,{\tilde {\bf
r}}_{{\rm O}_{1} {\rm O}_{2} } )} \!\right)}{\partial {\tilde
{r}}_{{\rm H}_{1} } } \right|_{\tilde {r}'_{{\rm H}_{1} } }
\!\!\!\!=0.\label{eq17}
\end{equation}
It should be noted that Eq.~(\ref{eq17}) gives rise to only an
insignificant shift $\Delta \tilde{r}_{\mathrm{H}_{1}}$ in the
equilibrium position of hydrogen $\mathrm{H}_{1}^{+}$.\,\,Supposing
that
\[
{{\tilde {r}}'}_{{\rm H}_{1} } ={\tilde {r}}_{{\rm H}_{1} } +\Delta
{\tilde {r}}_{{\rm H}_{1} } ,\,\,\,\vert \Delta {\tilde {r}}_{{\rm
H}_{1} } \vert \ll {\tilde {r}}_{{\rm H}_{1} },
\]
$\Delta \tilde{r}_{\mathrm{H}_{1}}$ can be determined with the help
of a simpler equation,
%18
\begin{equation}
\label{eq18} \left. \tilde {K}_{rr}^{(1)} \!\Delta {\tilde
{r}}_{{\rm H}_{1} } \!\!+\!\! {\boldsymbol{\nabla }}_{{\tilde
{r}}_{{\rm H}_{1} } } \!\tilde {\Phi }_{\rm Int} ({\tilde {\bf
r}}_{{\rm O}_{1} {\rm H}_{3} } ,{\tilde {\bf r}}_{{\rm O}_{1} {\rm
H}_{4} } ,{\tilde {\bf r}}_{{\rm O}_{1} {\rm O}_{2} } \!) \right|_
{\tilde {r}_{{\rm H}_{1}}
 =\tilde
{r}_{{\rm H}_{1} }' }\!\!\!=0,
\end{equation}
where $\tilde{K}_{rr}^{1}$ is the coefficient of elasticity for the bond
between the hydrogen and the oxygen in the monomer.

Equation (\ref{eq18}) acquires a very simple form in the case where
the distance $\tilde{r}_{\mathrm{O}_{1}\mathrm{O}_{2}}$ between the
oxygens in the dimer largely exceeds the diameter of a water
molecule.\,\,In this case, the potential
$\tilde{\Phi}_{\mathrm{Int}}(\mathbf{{\tilde{r}}}_{\mathrm{O}_{1}\mathrm{H}_{1}},\mathbf{{\tilde{r}}}_{\mathrm{O}_{1}\mathrm{H}_{2}},\mathbf{{%
\tilde{r}}}_{\mathrm{O}_{1}\mathrm{O}_{2}})$ looks like
%19
\[ \tilde {\Phi }_{\rm Int} (\tilde {\bf r}_{{\rm O}_{1}
{\rm H}_{1} } ,\tilde {\bf r}_{{\rm O}_{1} {\rm H}_{2} } ,\tilde
{\bf r}_{{\rm O}_{1} {\rm O}_{2} }\! )= q_i\Bigl[ \tilde {\Phi
}_{\rm Dq} (\tilde {\bf r}_{{\rm H}_{1} } ,...)\,+\]\vspace*{-7mm}
\begin{equation}
\label{eq19}  +\,\tilde {\Phi }_{\rm Dq} ({\tilde {\bf r}}_{{\rm
H}_{2} } ,...)-2\tilde {\Phi }_{\rm Dq} ({\tilde {\bf r}}_{{\rm
O}_{1} {\rm O}_{2} } ,...) \Bigr]\!,
\end{equation}
where \begin{equation*}
\tilde{\Phi}_{\rm Dq}(q_{i}\mathrm{,}\mathbf{{\tilde{d}}}_{\mathrm{w2}})=\frac{%
q_{i}\left(
{\mathbf{{\tilde{d}}}_{\mathrm{w2}}\mathbf{{\tilde{r}}}_{\mathrm{O}_{1}\mathrm{O}_{2}}}\right) }{\tilde{r}_{\mathrm{O}_{1}\mathrm{O}_{2}}^{3}}
\end{equation*}
is the potential of the charge-dipole interaction of hydrogens and
oxygen of molecule~1 with the electric field of molecule~2, which
can be approximated as the field of a dipole.\,\,In further
calculations, we used two approximations associated with the fact
that the hydrogen is subjected not only to electric forces, but also
to the repulsion forces from the oxygen electron shell.\,\,First, we
adopt that only the hydrogen position
$\mathbf{{\tilde{r}}}_{\mathrm{H}_{1}}$ is changed.\,\,In this case,
Eqs.~(\ref{eq18}) and (\ref{eq19}) yield
\[
\Delta {\tilde {\bf r}}_{{\rm H}_{1} } =-\frac{1}{\tilde
{K}_{rr}^{(1)} }\frac{{q}_{i} }{\vert {\tilde {\bf r}}_{{\rm O}_{1}
{\rm O}_{2} } +{\tilde {\bf r}}_{{\rm H}_{1} } \vert ^5}\Bigl[
{\tilde {\bf d}}_{{w2}} \vert {\tilde {\bf r}}_{{\rm O}_{1} {\rm
O}_{2} } +{\tilde {\bf r}}_{{\rm H}_{1} } \vert ^2
\,-\]\vspace*{-5mm}
\[-\,3\left(\! {\tilde {\bf d}}_{{w2}}
\left( {\tilde {\bf r}}_{{\rm O}_{1} {\rm O}_{2} } +{\tilde {\bf
r}}_{{\rm H}_{1} }  \right) \!\right)\left( {\tilde {\bf r}}_{{\rm
O}_{1} {\rm O}_{2} } +{\tilde {\bf r}}_{{\rm H}_{1} }  \right)
\Bigr]\!.
\]
With a satisfactory accuracy, this formula can be rewritten in the scalar
form
%20
\begin{equation}
\label{eq20} \Delta {\tilde {r}}_{{\rm H}_{1} }
=\frac{2}{K_{rr}^{(1)} }\frac{{q}_{i} {\tilde {d}}_{{w2}} \cos
\theta }{\vert {\tilde {\bf r}}_{{\rm O}_{1} {\rm O}_{2} } -{\tilde
{\bf r}}_{{\rm H}_{1} } \vert ^3},
\end{equation}

%Fig. 3
\begin{figure}%
\vskip1mm
\includegraphics[width=\column]{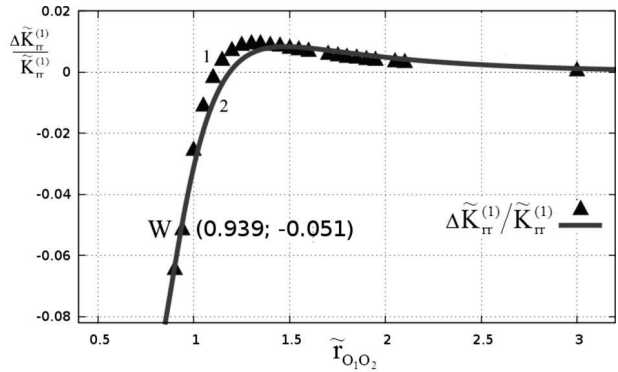}
\vskip-3mm\caption{Dependences of the relative force constant change
on the distance $\tilde{r}_{\mathrm{O}_{1}\mathrm{O}_{2}}$ in the
standard dimer calculated making allowance for the mutual adjustment
of molecular orientations (triangles,~{\it 1}) and for the
orientation of molecules as in Fig.~2 (solid curve,~{\it 2})
}%\vspace*{-1mm}
\end{figure}

%Tabl.3
\begin{table}[t]
\vskip5mm \noindent\caption{ Force constants for water
molecules}\vskip3mm\tabcolsep13.1pt

\noindent{\footnotesize\begin{tabular}{|l|c|c|c|}
  \hline
  \multicolumn{1}{|c}{\rule{0pt}{5mm}Constants} &
  \multicolumn{1}{|c}{$\frac {\partial^2 \tilde {\Phi}}{\partial \tilde {\rm r}_1 ^2}$ } &
  \multicolumn{1}{|c}{$\frac {\partial^2 \tilde {\Phi}}{\partial \theta^2}$ } &
  \multicolumn{1}{|c|}{$\frac {\partial^2 \tilde {\Phi}}{\partial \tilde {\rm r}_1 \partial \theta}$} \\[3mm]
  \hline \rule{0pt}{5mm}GSD model & 256.98 &182.15&33.439 \\
  Experiment [2] & 256.98 &190.64&33.439 \\[2mm]
  \hline
\end{tabular}}
\end{table}

%Tabl.4
\begin{table}[h!]
\vskip3mm \noindent\caption{Frequencies of symmetric\\ valence
vibrations of hydrogens in a water\\ molecule in vapor, liquid, and
ice }\vskip3mm\tabcolsep8.2pt

\noindent{\footnotesize\begin{tabular}{|l|c|c|c|}
  \hline \multicolumn{1}{|c}
{\raisebox{-3mm}[0pt][0pt]{Source}} & \multicolumn{3}{|c|}{\rule{0pt}{5mm}Frequencies}\\[1.5mm]%
\cline{2-4} \multicolumn{1}{|c}{}& \multicolumn{1}{|c}{$\nu_v$,
cm$^{-1}$}& \multicolumn{1}{|c}{$\nu_w$, cm$^{-1}$}&
\multicolumn{1}{|c|}{\rule{0pt}{4mm}$\nu_{\rm Ice}$, cm$^{-1}$}\\[1.5mm]%
\hline
\rule{0pt}{5mm}Experiment [1] & 3657\,~\, & 3490&3200 \\
  Experiment [8]& 3656.7 & 3280& \\
  MSN-FP [8] & 3656\,~\, & 3251& \\
  SPC-FP [8] &  & 3875& \\[2mm]
  \hline
\end{tabular}}\vspace*{-4mm}
\end{table}

\noindent where $F_1 =\frac{{q}_{i} {\tilde {d}}_{{w2}} \cos \theta
}{\vert {\tilde {\bf r}}_{{\rm O}_{1} {\rm O}_{2} } -{\tilde {\bf
r}}_{{\rm H}_{1} } \vert ^3}$ is the force that acts on the hydrogen
with the coordinate $\mathbf{{\tilde
{r}}}_{\mathrm{H}_{1}}$.\,\,Second, we took into account that the
electric field of the neighbor molecule also acts on the oxygen and
the hydrogen connected with it in molecule\,\,1.\,\,The approximate
expression for the force acting on the oxygen of molecule~2 looks
like $F_2 \approx -\frac{{q}_{i} {\tilde {d}}_{{w2}} \cos \theta
}{{\tilde {r}}_{{\rm O}_{1} {\rm O}_{2} }^3 }$. The sum $\delta
F_{1}$ of the forces $F_{1}$ and $F_{2}$ equals\vspace*{-1.5mm}
\[
\delta F_1 \approx 3\frac{{\tilde {d}}_{{\rm H}_{1} } {\tilde
{d}}_{{w2}} \cos \theta }{{\tilde {r}}_{{\rm O}_{1} {\rm O}_{2} }^4
},
\]
This is the force that stretches the vector
$\mathbf{{\tilde{r}}}_{\mathrm{H}_{1}}$.\,\,As a result, we obtain
the expression\vspace*{-1.5mm}
%21
\begin{equation}
\label{eq21} \delta {\tilde {r}}_{{\rm H}_{1} } \approx
\frac{3}{\tilde {K}_{rr}^{(1)} }\frac{{\tilde {d}}_{{\rm H}_{1} }
{\tilde {d}}_{{w2}} \cos \theta }{{\tilde {r}}_{{\rm O}_{1} {\rm
O}_{2} }^4 },
\end{equation}
which differs from Eq.\,\,(\ref{eq20}) by its functional dependence
on the distance between the oxygens
($\delta\tilde{r}_{\mathrm{H}_{1}}/\Delta
\tilde{r}_{\mathrm{H}_{1}}\sim3\tilde{r}_{\mathrm{H}_{1}}/\tilde
{r}_{\mathrm{O}_{1}\mathrm{O}_{2}}$).\,\,However, at distances of
several diameters of a water molecule, they are close by magnitude.

Taking into account that, by the order of magnitude, the elastic
constant $\tilde{K}_{rr}^{1}$ can be evaluated as
$\tilde{K}_{rr}^{1}\approx
q_{\mathrm{i}}^{2}/\tilde{r}_{\mathrm{H}}^{\mathrm{3}}$ in the
framework of the electrostatic model, we obtain the estimate for the
ratio $\delta\tilde
{r}_{\mathrm{H}_{1}}/\tilde{r}_{\mathrm{H}_{1}}$:
\[ {\delta \tilde
{r}}_{{\rm H}_{1} } {/\tilde {r}}_{{\rm H}_{1} } {\sim (\tilde
{r}}_{{\rm H}_{1} } /{\tilde {r}}_{{\rm O}_{1} {\rm O}_{2} }
)^4\leqslant 0{.}02.
\]
Therefore, the
shift can be neglected practically at all distances between the
oxygens in the water dimer.

The values of force constants for the water molecule corresponding
to the GSD potential are quoted in Table\,\,3.\,\,For comparison,
Table\,\,3 also contains the relevant values determined
experimentally.

The value of force constant $\tilde{K}_{rr}^{1}$ for symmetric
valence vibrations in the equilibrium dimer configuration shown in
Fig.~2 was calculated using formula~(\ref{eq15}).\,\,The relative
variation of the constant,
${\Delta\tilde{K}_{rr}^{1}/\tilde{K}_{rr}^{1}}$, as a function of
the distance between the oxygens in the dimer is plotted in Fig.~3.
Point W in the plot with the coordinates $(0.939, -0.051)$
corresponds to the distance
$r_{\mathrm{O}_{1}\mathrm{O}_{2}}=2.8~${\AA\ }between the oxygens,
which is typical of liquid water near the ternary point.\,\,The
relative change of the constant of symmetric valence vibrations
equals ${\Delta\tilde{K}_{rr}^{1}/\tilde{K}_{rr}^{1}}=$ $=-0.05$ at
this point.\,\,Hence, the values of force constant for valence
vibrations at the distance corresponding to the equilibrium state of
a dimer and the distance between water molecules in liquid water
differ from each other by 5\%.

\section{Discussion of the Results}

The frequency shift of hydrogen valence vibrations in a water
molecule depends on the phase state of water and can reaches a
magnitude of several hundreds of inverse centimeters
(Table~4).\,\,In this work, we suppose that the main contribution to
the experimentally observed magnitude of frequency shift is made by
the electrostatic forces connected with the multipole moments of
water molecules.

The basic result of our research consists in that the electrostatic
forces really give rise to the frequency shift values that agree
with experimental data by both the sign and the order of
magnitude.\,\,For instance, according to Eq.~(\ref{eq19}), the
frequency shift for valence vibrations equals
\[
{\Delta \omega }\approx \frac{{1}}{{1}}{\omega }_{0} \frac{{\Delta
\tilde {K}}_{{rr}}^{(1)} }{{\tilde {K}}_{{rr}}^{(1)} },
\]
where $\omega_{0}\approx3657$~\textrm{cm}$^{-1}$ is the vibration
frequency for an isolated water molecule.\,\,The relative increment
of the elastic constant equals
${\Delta\tilde{K}_{rr}^{1}/\tilde{K}_{rr}^{1}}=-0.05$ at
$r_{\mathrm{O}_{1}\mathrm{O}_{2}}=2.8~${\AA , }i.e.
$\Delta\omega\approx-0.025\omega _{0}=-91.43$~\textrm{cm}$^{-1}$.
Hence, the frequency shift sign for the dimer coincides with the
shift signs in liquid water and ice.\,\,The shear moduli are
identical by the order of magnitude, but, nevertheless, they are
appreciably different by the value.\,\,This circumstance has a
simple qualitative interpretation.\,\,The total electric field
acting on a water molecule in the liquid is, on the average, a
little stronger than that acting from the neighbor molecule in the
dimer.\,\,An insignificant increment of the electric field strength
in the case of the liquid is associated with a weakly ordered
arrangement of the centers-of-mass of the nearest neighbor molecules
and their orientations.\,\,As a consequence, according to the
superposition principle, only a weak enhancement of the electric
field in the molecule takes place.\,\,The situation in ice is
totally opposite.

Let us discuss a change of the elastic constant in the standard
dimer (see Fig.~2) at
$r_{\mathrm{O}_{1}\mathrm{O}_{2}}=2.85~\mathrm{\mathring{A}}$.\,\,It
is at this distance that the oxygens of water molecules are arranged
in the argon matrix.\,\,According to our calculations, the relative
increment of the elastic constant,
${\Delta\tilde{K}_{rr}^{1}/\tilde{K}_{rr}^{1}}$, in the concerned
dimer configuration amounts to $-0.0443$ at the fixed orientation of
molecules as in the standard dimer and to $-0.0432$ if the molecules
are allowed to adjust their orientations.\,\,Those variations in the
elastic constant stimulate changes in the frequency of valence
vibrations: to $3576$~\textrm{cm}$^{-1}$ in the former case and to
$3578$~\textrm{cm}$^{-1}$ in the latter one.\,\,It should be noted
that the relatively small value of orientational contribution is
explained by the fact that the configuration of the system is in the
interval where the energy of the repulsion between the molecules
monotonically decreases.\,\,The frequency corresponding to
experiments in the argon matrix \cite{7} equals
3574~\textrm{cm}$^{-1}$.\,\,Therefore, we believe that it is
possible to say about the complete coincidence of theoretical and
experimental results.\,\,From our viewpoint, this is a sound
argument in favor of the electrostatic nature of the frequency shift
of valence vibrations.

It is very important that the explanation of the frequency shifts
different by the magnitude is essentially based on the superposition
principle, the application of which to sharply directed and
saturated irreducible hydrogen bonds is impossible.\,\,For a more
complete substantiation of this fact, we intend to consider the
frequency shifts of valence vibrations in ice and liquid water in a
separate work.

Not less important is the fact that, in rarefied vapor, one should
expect a positive frequency shift, which directly follows from the
behavior of the quantity
${\Delta\tilde{K}_{rr}^{1}/\tilde{K}_{rr}^{1}}$ depicted in
Fig.\,\,3.\,\,This fact is also qualitatively supported by
experimental data \cite{29} on the IR absorption in relatively
rarefied water vapor.

\vskip3mm {\it We would like to express our deep respect to the late
Galyna Oleksandrivna Puchkovska, who stimulated this work about 20
years ago.\,\,A delay in the solution of the formulated problem was
connected with the absence of a suitable interaction potential
between soft water molecules.\,\,Professor V.E.\,Pogorelov, who
studied the shifts of valence vibrations in alcohols of the methanol
series, permanently induced us to the solution of the formulated
problem.\,\,Therefore, we consider Prof.\,G.O.\,Puchkovska and
Prof.\,V.E.\,Pogorelov to be the principal authors of this
paper.\,\,It is especially pleasant to mark this circumstance in
connection with the 80-th anniversary of the birthday of
unforgettable Galyna Oleksandrivna Puchkovska, which was in June
this year.\,\,We are also sincerely grateful to Academician
L.A.\,Bulavin for the comprehensive discussion of the results of
this work at its various stages.}

\vspace*{1mm}
\rezume{%
І.В.\,Жиганюк, М.П.\,Маломуж}{ПРИРОДА ЗСУВУ ЧАСТОТИ ВАЛЕНТНИХ\\
КОЛИВАНЬ ГІДРОГЕНІВ У МОЛЕКУЛАХ ВОДИ} {Робота присвячена дослідженню
фізичної природи зсуву частот валентних коливань гідрогенів молекул
води внаслідок її взаємодії з сусідніми молекулами. Приймається, що
домінуючий внесок у міжмолекулярну взаємодію вноситься силами
електростатичної природи, пов'язаними з існуванням мультипольних
моментів молекул води. Розраховано величину зсуву частоти у випадку,
коли дві сусідні молекули води утворюють димер. Отриманий результат
якісно добре узгоджується з величиною зсувів частот, які
спостерігаються у парі, льоді та рідкій воді, а також у водних
розчинах спиртів [1--4]. Це свідчить про те, що водневі зв'язки, за
допомогою яких намагаються відтворити специфіку міжмолекулярної
взаємодії у воді, а також її макроскопічні властивості, формуються
домінуючим чином силами електростатичної природи.}

\end{document}